\documentclass[sigconf]{acmart}

\usepackage{tikz}
\usetikzlibrary{shapes.geometric, arrows.meta, positioning, backgrounds, patterns}
\usepackage{pifont}
\usepackage{colortbl}
\usepackage{booktabs}
\usepackage{listings}
\usepackage{dblfloatfix}
\usepackage{subcaption}
\usepackage{multirow}
\usepackage{makecell}
\usepackage{seqsplit}
\definecolor{clrEager}{HTML}{D1D5DB}
\definecolor{clrFusion}{HTML}{93C5FD}
\definecolor{clrTiling}{HTML}{6EE7B7}
\definecolor{clrCutlass}{HTML}{FCD34D}
\definecolor{clrArch}{HTML}{F87171}
\definecolor{clrGap}{HTML}{DC2626}
\definecolor{clrOurs}{HTML}{2563EB}

\AtBeginDocument{%
  }

\setcopyright{acmlicensed}
\copyrightyear{2026}
\acmYear{2026}
\acmDOI{XXXXXXX.XXXXXXX}
\acmConference[]{}{}{}
\acmISBN{978-1-4503-XXXX-X/26/09}

\begin{document}

\title{FACT: Compositional Kernel Synthesis with a Three-Stage Agentic Workflow}

\author{Sina Heidari}
\email{sinaheidari@vt.edu}

\affiliation{%
  \institution{Virginia Tech}
  \city{Blacksburg}
  \state{Virginia}
  \country{USA}
}

\author{Dimitrios S. Nikolopoulos}
\email{dsn@vt.edu}
\affiliation{%
  \institution{Virginia Tech}
  \city{Blacksburg}
  \state{Virginia}
  \country{USA}
}

\renewcommand{\shortauthors}{Heidari and Nikolopoulos}

    \begin{abstract}
Deep learning compilers and vendor libraries deliver strong baseline performance but their performance is bounded by finite, engineer-curated catalogs. When these omit needed optimizations, practitioners substitute hand-written CUDA or CUTLASS, demanding expertise in GPU microarchitecture and C++ template metaprogramming. Recent LLM-based agents target kernel generation in raw CUDA, forcing rediscovery of optimizations already encoded in mature libraries. We present FACT (Framework for Agentic CUTLASS Transpilation), a three-stage agent-driven workflow optimizing PyTorch modules through multi-pattern composition while grounding synthesis in CUTLASS C++. Pattern discovery inspects the traced graph, matches subgraphs to optimization rules, retrieves vetted examples, and outputs prioritized patterns. Pattern realization implements each pattern as a CUTLASS kernel, verifies, and auto-tunes. Pattern composition assembles extensions into an optimized module for benchmarking. We evaluate the workflow on KernelBench across NVIDIA A100 and H100 GPUs. On Level 1 GEMM problems (square, batched, large-K matrix multiply), auto-tuned CUTLASS kernels achieve $1.06\times$--$1.18\times$ speedups on A100 and $0.84\times$--$1.80\times$ performance variations on H100 over cuBLAS. On Level 3 transformer blocks against PyTorch eager baseline, FACT achieves $2.03\times$ speedup on MiniGPT (vs. Inductor: $1.89\times$, TensorRT: $1.85\times$) and $1.41\times$ on Llama 3 8B (vs. Inductor: $1.17\times$, TensorRT: $1.18\times$). Our framework couples agentic graph-level pattern discovery with architecture-specific auto-tuning and a dynamic pattern registry, offering a practical path from traced PyTorch modules to deployable kernels.
  \end{abstract}

\ccsdesc[500]{Computing methodologies~Parallel computing methodologies}
\ccsdesc[500]{Computing methodologies~Machine learning}
\ccsdesc[300]{Software and its engineering}

\keywords{CUTLASS Kernel Synthesis, Architecture-aware Optimization, Agentic Code Generation, GPU Kernel Optimization, Multi-pattern Composition, Large Language Models}

\maketitle

\section{Introduction}
\label{sec:introduction}

Deep learning compilers and SDKs such as TVM~\cite{chen2018tvm}, Ansor~\cite{zheng2020ansor}, PyTorch Inductor~\cite{ansel2024pytorch2}, and NVIDIA TensorRT~\cite{nvidia_tensorrt} have made significant strides in automating the compilation of tensor programs. However, their ability to generate high-performance GPU kernels is fundamentally constrained by the optimization patterns that their developers have codified. When a new model architecture or a new GPU generation is released, developing optimized kernels that exploit its novel hardware features requires significant engineering effort, and integrating these optimizations into compiler backends adds further delay before they become available to practitioners. Even when optimized kernels are eventually integrated into a compiler---as CUrator~\cite{curator_cgo2025} does by incorporating CUTLASS~\cite{nvidia_cutlass_docs} into TVM's Bring-Your-Own-Codegen (BYOC) pipeline~\cite{byoc_tvm}---the compiler relies on a fixed, developer-specified pattern registry for graph node matching and fusion, which cannot discover or add new optimization patterns without manual intervention.

The alternative to waiting for compiler support is to hand-optimize kernels using low-level languages such as CUDA or vendor-specific template libraries such as CUTLASS. Expert GPU programmers can write custom kernels that exploit architecture-specific features and integrate them into a PyTorch computation graph via custom extensions. This approach routinely achieves performance competitive with or exceeding compiler output, but it demands deep expertise in GPU microarchitecture and months of engineering effort per kernel.

Over the past two years, large language model (LLM)-based agents have demonstrated remarkable capabilities in code generation~\cite{nijkamp2023codegen}, program synthesis~\cite{zelikman2023parsel}, and multi-step reasoning~\cite{wei2023chain}. These capabilities have been applied directly to GPU kernel optimization: CUDA Agent~\cite{cuda_agent_arxiv2026} scales agentic reinforcement learning to generate high-performance CUDA kernels; KernelBlaster~\cite{kernelblaster_arxiv2026} uses memory-augmented in-context RL to transfer optimization knowledge across tasks; Astra~\cite{astra_arxiv2025} employs a multi-agent workflow to iteratively transform existing kernels using profiling-driven feedback; and StitchCUDA~\cite{stitchcuda_arxiv2026} targets end-to-end GPU programs with a planner/coder/verifier decomposition.

However, these approaches share a common limitation: they synthesize kernels in raw CUDA, which forces them to rediscover---in generated source code---optimizations that mature libraries have already encapsulated as vetted, architecture-parameterized templates. Despite the promise of agentic kernel optimization, we are not aware of any prior framework that combines library-grounded synthesis with a dynamic pattern registry---addressing both the static registry limitation of compiler-based approaches and the raw-code generation focus of prior agentic systems.

 We propose FACT (Framework for Agentic CUTLASS Transpila-
tion), a framework that combines multi-pattern composition with library-grounded synthesis through a dynamic pattern registry that grows as the agent discovers new patterns---where each pattern corresponds to a subgraph identifiable by an optimization rule (e.g., GEMM, FMHA), data type, and target GPU architecture---unlike static registries, this enables accumulation across workloads. FACT performs architecture-specific synthesis by matching patterns to the target GPU's capabilities. It also enables architecture-specific auto-tuning where the agent automatically infers which parameters to sweep at each level of CUTLASS's hierarchical API.

FACT realizes this through a three-stage agentic workflow. (1)~Pattern Discovery: an LLM agent analyzes the PyTorch computation graph and proposes optimization patterns by retrieving vetted examples from an architecture-specific kernel repository. (2)~Pattern Realization: each pattern is implemented as a CUTLASS kernel, verified for correctness, auto-tuned, and added to the dynamic pattern registry. (3)~Pattern Composition: realized kernels are composed with the original computation graph for end-to-end benchmarking.

Both synthesis and auto-tuning navigate CUTLASS's three-level hierarchy. At the tile level, synthesis configures Tensor Core primitives and tile shapes. At the kernel level, synthesis selects pipelining strategies appropriate for the target architecture. At the grid level, synthesis chooses scheduling policies that balance load across thread blocks. These levels must be addressed coherently—\seqsplit{incompatible} choices across levels cause failures. Auto-tuning sweeps parameters at each level to find optimal configurations.

The contributions of this paper are as follows:
\begin{itemize}
  \item \textbf{Library-grounded agentic kernel synthesis with a dynamic pattern registry.} We ground the agent's synthesis in CUTLASS C++ templates rather than raw CUDA, enabling reuse of architecture-parameterized patterns that accumulate across workloads.
  \item \textbf{Architecture-specific auto-tuning where the agent infers relevant parameters based on the target architecture and kernel type.} For Ampere multistage pipelines, the agent sweeps tile shapes and pipeline stages; for Hopper warp-specialized kernels, it sweeps tile shapes, kernel schedules and cluster shapes.
  \item \textbf{KernelBench-style evaluation on representative problems spanning CUTLASS's capabilities.} On Level 1 GEMM problems (square, batched, and large-K matrix multiply), auto-tuned CUTLASS kernels achieve $1.06\times$--$1.18\times$ speedups on A100 and $0.84\times$--$1.80\times$ performance variations on H100 over cuBLAS. On Level 3 transformer blocks, FACT achieves $2.03\times$ speedup on MiniGPT and $1.41\times$ on Llama 3 8B, consistently outperforming compiler baselines.
  \item \textbf{Open-source artifacts and reproducibility.} We release the pattern registry and synthesized kernels (via GLM~4.7) developed through the
   framework, along with the LLM prompts for our three-stage agentic workflow, enabling              
  reproducibility and community extensions. Available at \href{https://github.com/Project-FACT/FACT}{github.com/Project-FACT/FACT}.
\end{itemize}

\section{Background \& Motivation}

\subsection{KernelBench Evaluation Framework}

KernelBench~\cite{kernelbench} introduces a standardized benchmark for evaluating LLM agents on GPU kernel generation. The framework adopts a two-stage evaluation: (1) \emph{correctness}, where synthesized kernels are compiled and compared against PyTorch references, and (2) \emph{performance}, where correctly-executing kernels are benchmarked and compared to baselines. KernelBench defines three difficulty tiers: \textbf{Level~1} (single operators like GEMM), \textbf{Level~2} (fused operators), and \textbf{Level~3} (complex blocks from real models). The optimization problem is formulated as \emph{transpiling} a PyTorch computation graph into functionally equivalent kernel implementations that replace baseline subgraphs in the execution trace, closely mirroring real-world kernel optimization workflows.

\textbf{CUTLASS.} CUTLASS~\cite{nvidia_cutlass_docs} is a C++ template library for efficient GPU kernel synthesis and optimization. It provides a modular, composable architecture spanning three levels of abstraction: tile-level primitives for Tensor Core operations (e.g., WGMMA for Hopper's matrix-multiply-accumulate units), kernel-level pipelining strategies (warp-specialized pipelines for Hopper, multistage pipelines with cp.async for Ampere), and grid-level work scheduling (dual mode vs.\ warp-specialized modes for load balancing and synchronization). This multi-level design enables architecture-specific kernel customization while maintaining a consistent programming model, making CUTLASS an ideal foundation for automated kernel synthesis.

We leverage CUTLASS to implement fused multi-head attention (FMHA) kernels inspired by FlashAttention~\cite{dao2022flashattention}, which reduces memory I/O by tiling over the attention computation and keeping intermediate results in on-chip SRAM rather than writing to and reading from HBM.

\section{Task Definition}
\label{sec:task_definition}

We adopt a task formulation extending KernelBench's transpilation paradigm with two key contributions. First, we target \emph{multi-pattern composition} rather than isolated operators: the agent operates over the complete computation graph to discover and optimize multiple subgraph patterns simultaneously. Second, we introduce a \emph{dynamic pattern registry} indexed by a pattern table $\mathcal{T}$ that maps optimization rules ($r$), data types ($\tau$), and target architectures ($\alpha$) to CUTLASS kernel implementations. This design enables efficient retrieval of architecture-specific kernels and accumulation of optimization knowledge across workloads.

The agent's objective is to discover optimization patterns $\mathcal{P} = \{p_1, p_2, \ldots, p_n\}$ where each pattern $p_i$ corresponds to a computation subgraph that can be replaced with an optimized kernel. Each pattern is associated with a \emph{rule}---such as GEMM, FMHA (Fused Multi-Head Attention), or epilogue fusion---that determines the kernel synthesis template. The pattern table $\mathcal{T}$ dynamically grows as new patterns are synthesized: when the agent discovers a correct kernel implementation, it is added to the registry indexed by $(r, \tau, \alpha)$, enabling future retrieval and reuse across models.

\begin{figure}[t]
\centering
\begin{tikzpicture}[
    node distance=0.8cm and 0.5cm,
    box/.style={rectangle, draw=black!60, fill=white, thick, minimum width=2cm, minimum height=0.7cm, align=center, font=\small},
    arrow/.style={-Stealth, thick, color=clrOurs},
    smallbox/.style={rectangle, draw=none, fill=none, align=center, font=\scriptsize}
]
\node[box, fill=clrEager!30] (graph) {Comp. Graph\\$\mathcal{G}$};
\node[smallbox, below=0.1cm of graph, xshift=-0.7cm, color=gray] (graphlabel) {PyTorch model};

\node[box, fill=clrOurs!30, below=of graph] (agent) {Agent};
\node[smallbox, below=0.1cm of agent, xshift=-0.8cm, color=gray] (agentlabel) {Pattern Discovery};

\node[box, fill=clrFusion!30, below=of agent] (patterns) {Patterns\\$\mathcal{P}$};
\node[smallbox, below=0.1cm of patterns, xshift=-0.5cm, color=gray] (patternslabel) {Subgraphs};

\draw[arrow] (graph) -- (agent);
\draw[arrow] (agent) -- (patterns);

\node[box, fill=clrCutlass!30, below=of patterns] (table) {Pattern Table\\$\mathcal{T}$};
\node[smallbox, below=0.1cm of table, xshift=-0.8cm, color=gray] (tablelabel) {Dynamic catalog};
\draw[arrow] (patterns) -- (table);

\node[box, fill=clrTiling!30, right=0.8cm of table, minimum width=1.5cm] (index) {Index\\$(r,\tau,\alpha)$};
\draw[arrow, dashed] (table) -- (index);

\node[smallbox, above=0.05cm of index, color=clrArch, align=left] (rule) {$r$: Rule\\(GEMM, FMHA)};
\node[smallbox, right=0.05cm of index.north east, color=clrArch, align=left, anchor=north west] (dtype) {$\tau$: Type\\(FP32, BF16)};
\node[smallbox, below=0.05cm of index, color=clrArch, align=left] (arch) {$\alpha$: Arch\\(Ampere, Hopper)};

\node[box, fill=clrGap!20, below=of table] (kernels) {CUTLASS\\Kernels};
\node[smallbox, below=0.1cm of kernels, color=gray] (kernelslabel) {Specialized impl.};
\draw[arrow] (table) -- (kernels);

\end{tikzpicture}
\caption{Pattern mapping hierarchy. The agent identifies optimization patterns from the computation graph, each associated with a rule (e.g., GEMM, FMHA). The pattern table $\mathcal{T}$ indexes kernels by a tuple $(r,\tau,\alpha)$ capturing rule, data type, and target architecture, enabling retrieval of workload-matched CUTLASS implementations.}
\label{fig:pattern_mapping}
\end{figure}
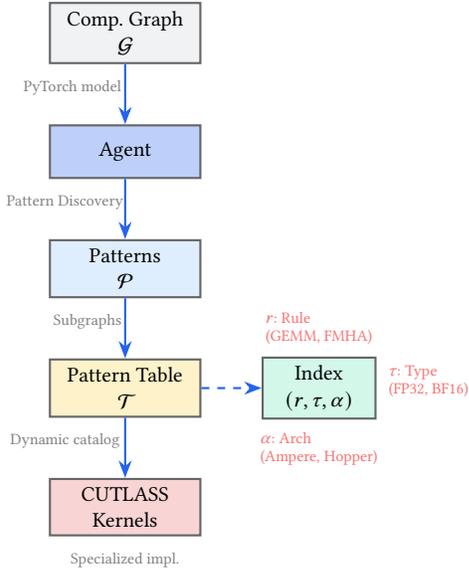

\section{Proposed Approach}

\begin{figure*}[!htbp]
\centering
\includegraphics[width=0.9\textwidth,height=0.4\textheight,keepaspectratio,trim=5 5 5 5,clip]{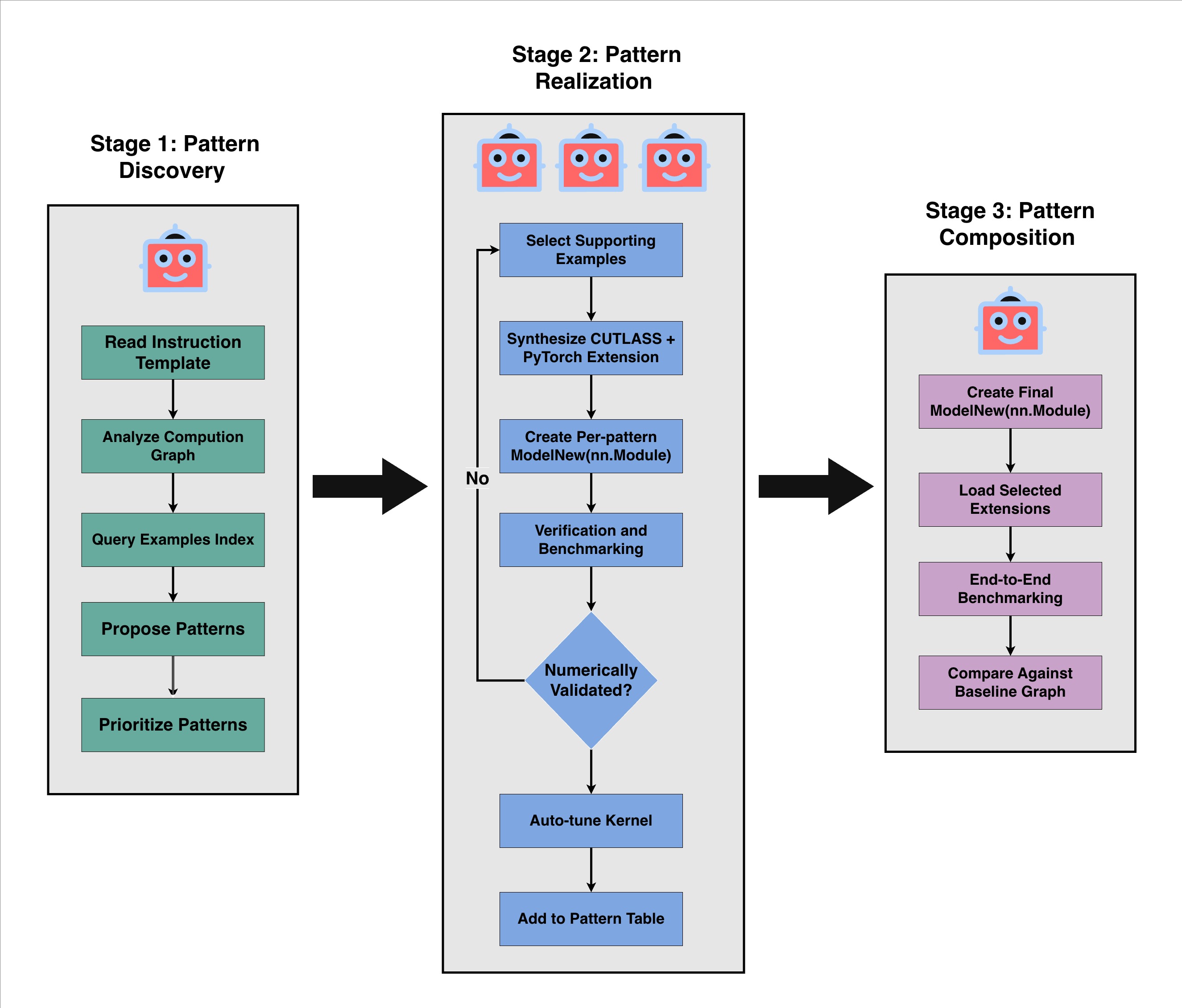}
\caption{Three-stage agentic workflow for whole-model kernel optimization. Stage 1 (Pattern Discovery) extracts and analyzes the computation graph to identify optimization opportunities. Stage 2 (Pattern Realization) synthesizes CUTLASS kernels for each pattern, verifying correctness and benchmarking performance before adding to the dynamic pattern registry. Stage 3 (Pattern Composition) assembles the optimized kernels into a deployable model with PyTorch extensions and measures end-to-end speedup. LLM-based actions enable systematic execution of complex optimization tasks across all stages.}
\label{fig:workflow}
\end{figure*}

Figure~\ref{fig:workflow} illustrates our three-stage agentic workflow for whole-model kernel optimization. The workflow maps directly to CUTLASS's API hierarchy~\cite{nvidia_cutlass_docs} and leverages LLM-based actions to orchestrate complex optimization tasks that would otherwise require manual intervention. Each stage decomposes into a sequence of well-defined actions---graph extraction and analysis, kernel synthesis and verification, composition and extension loading---that the agent executes systematically. By structuring optimization as a pipeline of discrete actions, we enable the agent to reason about each decision point independently while maintaining coherence across the entire workflow.

\subsection{Pattern Discovery}

The first stage begins with the baseline PyTorch model and executes five sequential actions to discover optimization opportunities.

\textbf{Action 1: Read instruction template.} The agent first reads an instruction template that defines the pattern discovery task. This template specifies the optimization objectives, the target GPU architecture, data type constraints, and examples of CUTLASS patterns to search for. The instruction grounds the agent's analysis in domain knowledge about CUTLASS capabilities.

\textbf{Action 2: Analyze computation graph.} Using PyTorch's tracing mechanism, the agent extracts the computation graph $\mathcal{G} = (V, E)$ as a sequence of tensor operations. This analysis preserves operator semantics, tensor shapes, and data type information. The agent then performs structural pattern matching to locate subgraphs that correspond to known CUTLASS patterns. The agent leverages PyTorch's tracing infrastructure—specifically \seqsplit{torch.jit.trace()} and \seqsplit{torch.fx.Tracer()}—to programmatically extract computation graphs from the target model, enabling systematic identification of subgraphs that match known CUTLASS patterns. During this analysis, the agent may not find exact matches in the examples index; however, it can retrieve multiple examples that, when combined, provide the necessary components to realize the target pattern.

\textbf{Action 3: Query examples index.} For each candidate subgraph, the agent retrieves relevant CUTLASS kernel implementations from an architecture-specific examples index organized by optimization rule, data type, and input shape. By grounding synthesis in proven implementations rather than generating from scratch, the agent improves reliability and performance. Table~\ref{tab:examples_by_level} summarizes available examples across KernelBench difficulty levels (Level~1: single operators; Level~2: fused operators; Level~3: complex transformer blocks) and GPU architectures.

\begin{table*}[t]
\centering
\small
\setlength{\tabcolsep}{4pt}
\begin{tabular}{p{0.15\textwidth} p{0.28\textwidth} p{0.28\textwidth} p{0.28\textwidth}}
\toprule
\textbf{Architecture} & \textbf{Level~1: Single Operator} & \textbf{Level~2: Fused Operators} & \textbf{Level~3: Complex Blocks} \\
\midrule
\textbf{Hopper} & Warp-Specialized GEMM & Epilogue Swizzle & FMHA \\
(SM90) & Topk-Softmax GEMM & Gather-Scatter Fusion & Grouped GEMM \\
& Sparse GEMM & GEMM-Permute & FP8 Grouped GEMM \\
& FP8 GEMM & & Mixed-DType Grouped GEMM \\
\midrule
\textbf{Ampere} & TF32 TensorOp GEMM & Operand-Reduction Fusion & Fused Multi-Head Attention \\
(SM80) & Sparse TensorOp GEMM & Fprop Mainloop Fusion & Grouped GEMM \\
& TensorOp Conv2dFprop & Wgrad Mainloop Fusion & GEMM-LayerNorm-GEMM Fusion \\
& FP64 Affine2 GEMM & GEMM-Softmax & Multi-GEMM IR Codegen \\
& TensorOp Group Conv & Gather-Scatter Fusion & Dual GEMM \\
\bottomrule
\end{tabular}
\caption{CUTLASS examples organized by KernelBench difficulty level and GPU architecture. Level~1: single linear algebra operators; Level~2: fused operators with epilogue or data movement fusion; Level~3: complex multi-operator patterns requiring interdependent scheduling decisions.}
\label{tab:examples_by_level}
\end{table*}

\textbf{Action 4: Propose patterns.} Based on the subgraph analysis and retrieved examples, the agent proposes a set of optimization patterns $\mathcal{P}_{\text{proposed}} = \{p_1, p_2, \ldots, p_m\}$. Each proposed pattern includes the subgraph to be optimized, the associated optimization rule (e.g., GEMM, FMHA), and references to the retrieved examples that guide kernel synthesis.

\textbf{Action 5: Prioritize patterns.} The agent prioritizes the proposed patterns based on expected performance impact and implementation complexity. This prioritization considers factors such as operator granularity, memory access patterns, and the quality of retrieved examples. The output is an ordered list of prioritized patterns $\mathcal{P} = \{p_1, p_2, \ldots, p_n\}$ that are passed to the next stage for kernel synthesis.

\subsection{Pattern Realization}

The second stage synthesizes concrete CUTLASS kernels for each prioritized pattern. For each pattern $p_i \in \mathcal{P}$, the agent executes an iterative workflow with six actions, including a feedback loop for correctness verification and an optional auto-tuning step.

\textbf{Action 1: Select supporting examples.} The agent queries the examples index to retrieve CUTLASS kernel implementations that serve as references for synthesizing the target pattern. These supporting examples are selected based on similarity to the target pattern in terms of optimization rule, data type, input shape, and GPU architecture. The agent may retrieve multiple examples that, when combined, provide the necessary components to realize the pattern.

\begin{figure}[h]
\centering
\includegraphics[width=0.48\textwidth, height=6cm]{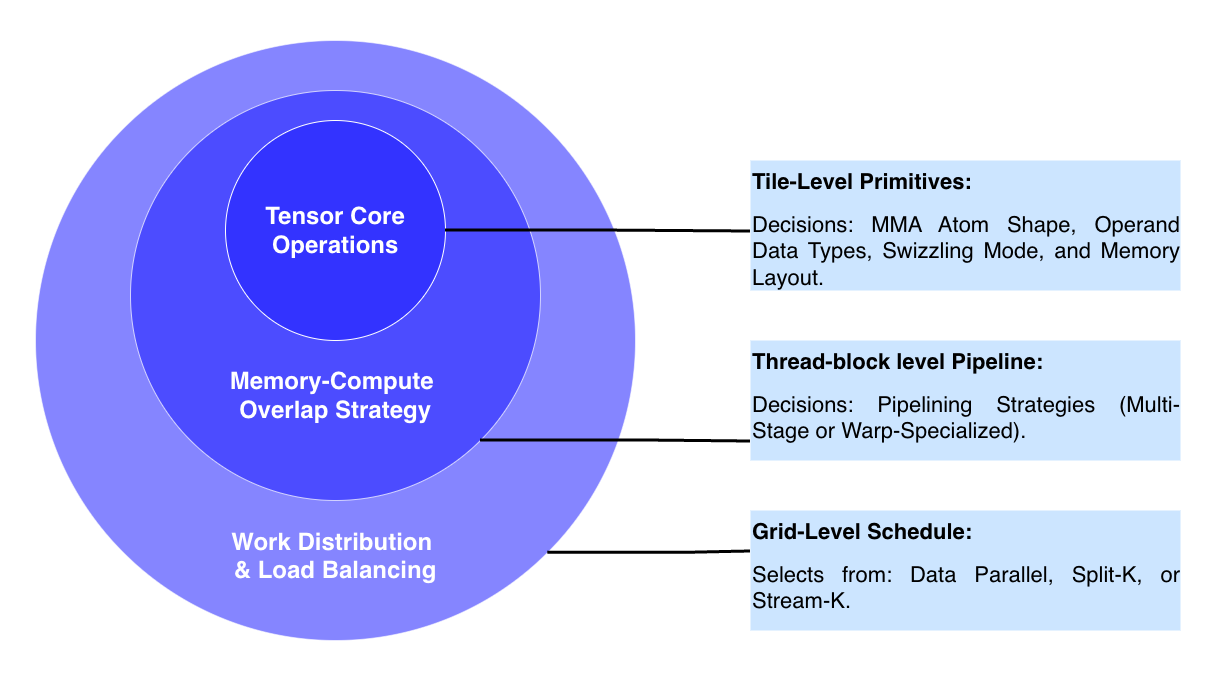}
\caption{Three-level hierarchy of CUTLASS kernel synthesis. From innermost to outermost: (1) Tile-level primitives; (2) Kernel-level pipelining; (3) Grid-level scheduling.}
\label{fig:cutlass_levels}
\end{figure}

\textbf{Action 2: Synthesize CUTLASS + PyTorch extension.} Guided by supporting examples, the agent synthesizes a CUTLASS kernel by configuring three API levels as shown in Figure~\ref{fig:cutlass_levels}: (1) tile-level primitives for Tensor Core operations, (2) kernel-level pipelining strategies appropriate for the target architecture (warp-specialized for Hopper, multistage pipelines for Ampere), and (3) grid-level work scheduling (Data Parallel, Split-K, or Stream-K). The synthesized kernel is wrapped as a PyTorch extension, compiled into a shared library, and registered as a custom operation.

\textbf{Action 3: Create per-pattern ModelNew(nn.Module).} The agent constructs a new PyTorch module, \texttt{ModelNew}, that integrates the synthesized CUTLASS extension. This module inherits from \texttt{nn.Module} and implements the \texttt{forward} method to invoke the custom CUTLASS operation. The module preserves the input/output interface of the original subgraph, enabling drop-in replacement within the computation graph.

\textbf{Action 4: Verification and benchmarking.} The agent executes \texttt{ModelNew} against a reference implementation using identical test inputs. For correctness verification, output tensors are compared element-wise to ensure functional equivalence. For performance benchmarking, the kernel is executed multiple times on the target GPU hardware, and CUDA events measure end-to-end execution time. The resulting speedup relative to the baseline PyTorch implementation is recorded.

At this point, the agent evaluates whether the pattern is \emph{accepted}. If the pattern passes both correctness verification and achieves satisfactory performance, the workflow proceeds to Action~5. If the pattern fails verification or performs poorly, the workflow loops back to Action~1: the agent re-selects supporting examples, inspects the synthesized code to identify correctness issues, and modifies the implementation to fix the problems. This iterative refinement continues until the pattern is accepted.

\textbf{Action 5: Auto-tune kernel.} After correctness verification, the agent performs auto-tuning to optimize performance. The search space is inferred from the CUTLASS API levels and target architecture rather than hardcoded. For Ampere multistage pipelines, the agent sweeps threadblock tile shapes, warp tile shapes, and pipeline stages. For Hopper warp-specialized kernels, the agent sweeps threadblock tile shapes, cluster shapes, and kernel schedules (cooperative and pingpong). Each configuration is compiled, executed with warmup and timed iterations, and the best-performing variant is selected for subsequent stages.

\textbf{Action 6: Add to pattern registry.} Once a pattern is accepted, the agent adds it to the dynamic pattern registry indexed by $\mathcal{T}$. The entry includes the pattern $p_i$, the synthesized CUTLASS kernel implementation, the auto-tuning results (if performed), the benchmark results (speedup metrics), and metadata about the supporting examples used. The entry is indexed by the tuple $(r, \tau, \alpha)$, enabling future queries to retrieve the kernel without re-synthesis. This action expands the optimization catalog, allowing the agent to accumulate knowledge across patterns and reuse kernels for similar workloads.

\subsection{Pattern Composition}

The third stage composes realized kernels into an optimized model. The agent constructs a new computation graph $\mathcal{G}'$ by replacing each subgraph $\mathcal{G}_i$ with a CUTLASS kernel call, preserving semantic equivalence. Each kernel is compiled as a PyTorch extension and loaded into the Python runtime. The agent then performs end-to-end benchmarking to validate that kernel optimizations translate to model-level speedup. The final output is a deployable PyTorch model with integrated CUTLASS kernels, and the pattern registry persists for future optimization sessions.

\section{Performance Evaluation}

\subsection{Experimental Setup}
\label{sec:experimental_setup}

\textbf{Notation.}
Throughout the evaluation, we use $(B, T, C)$ to denote tensor dimensions: $B$ is batch size, $T$ is sequence length, and $C$ is hidden dimension.

\textbf{Hardware.}
All experiments were conducted on two NVIDIA GPU systems: an A100-SXM4-80GB (Ampere architecture, SM80), which provides theoretical peak throughput of 156~TFLOPS in TF32 mode and 312~TFLOPS in FP16 mode~\cite{nvidia_a100_datasheet}, and an H100-SXM (Hopper architecture, SM90) with theoretical peak throughput of 989~TFLOPS in TF32 mode and 1,979~TFLOPS in FP16 mode~\cite{nvidia_hopper_whitepaper}.

\textbf{Benchmark problems.}
We evaluate on three KernelBench Level~1 problems drawn from the GEMM family (Problems~1, 3, and~6), each exercising a different grid-level scheduling policy: Data Parallel, Batched, and Stream-K, respectively. \textit{Problem~1} computes $C = A \times B$ for $4096 \times 4096$ matrices in FP32, exercising a standard square GEMM workload that maps directly to CUTLASS's TF32 tensor-op GEMM template with Data Parallel scheduling.
\textit{Problem~3} computes a batched matrix multiplication $C[b] = A[b] \times B[b]$ for $b \in [0, 128)$ with per-batch dimensions $512 \times 1024$ and $1024 \times 2048$, evaluating Batched GEMM via CUTLASS's \seqsplit{\texttt{kBatched}} mode.
\textit{Problem~6} computes $C = A \times B$ for $256 \times 524{,}288$ and $524{,}288 \times 256$ matrices, stressing the K dimension to evaluate Stream-K grid-level scheduling~\cite{osama2023streamk}.
Together, Problems~1, 3, and~6 span the three dominant scheduling regimes for GEMM on modern GPUs.

For multi-pattern workloads, we evaluate on two Level~3 transformer blocks:
(1) KernelBench problem~\seqsplit{\texttt{44\_MiniGPTBlock}}, a \seqsplit{MiniGPT-Style} block with LayerNorm, GELU activation, and dimensions $(128, 512, 768)$.
(2) A Llama~3 8B decoder block extracted from HuggingFace (not part of KernelBench), featuring RMSNorm, SwiGLU activation, Grouped-Query Attention (GQA), and dimensions $(16, 2048, 4096)$ with intermediate dimension 14,336. This block presents a more complex computation graph than \mbox{MiniGPT}.

\textbf{Software environment.}
Kernels were implemented using CUTLASS~3.8.0 (git commit~ea46e277), compiled with NVCC through PyTorch's \texttt{\seqsplit{torch.utils.cpp\_extension}} interface, and loaded as custom CUDA extensions.
Experiments on the A100 node were conducted on Ubuntu~24.04 with CUDA~12.4, PyTorch~2.4.1+cu124, and Torch-TensorRT~2.4.0.
Experiments on the H100 node were conducted with CUDA~12.8 and PyTorch~2.8.0.
Random inputs were generated using \seqsplit{\texttt{torch.manual.seed(42)}} for reproducibility.
For Level~1 representative GEMM problems, the baseline is PyTorch eager-mode execution with \seqsplit{torch.backends.cuda.matmul.allow\_tf32 = True}, which dispatches to cuBLAS's TF32 tensor core GEMM implementation.
For the Level~3 multi-pattern problems, we additionally evaluated compiler baselines: PyTorch Inductor (using torch.compile with backend=\seqsplit{\texttt{"inductor"}} and mode=\seqsplit{\texttt{"max-autotune"}}) and Torch-TensorRT (using \seqsplit{\texttt{torch.compile}} with backend=\seqsplit{\texttt{"torch\_tensorrt"}} and dynamic=False).
Correctness was verified using \seqsplit{\texttt{torch.allclose()}} with \seqsplit{\texttt{rtol=1e-3}} and \seqsplit{\texttt{atol=1e-5}}, comparing the CUTLASS kernel output against the PyTorch reference.

\textbf{Auto-tuning methodology.}
Auto-tuning is performed by the agent as Action~5 in the Pattern Realization stage (Section~4.2). The agent infers the search space from the CUTLASS test files for the target architecture and data type.
For Ampere (SM80) multistage pipelines, the agent sweeps threadblock tile shapes, warp tile shapes, and pipeline stages.
For Hopper (SM90) warp-specialized kernels, the agent sweeps threadblock tile shapes, cluster shapes, and kernel schedules (cooperative and pingpong).
Each configuration was compiled as a standalone shared library, executed with 5~warmup iterations followed by 20~timed iterations, and measured using \seqsplit{\texttt{time.perf\_counter}} with CUDA synchronization. All reported timings are averaged over 20 trials; variance was negligible (<1\% across configurations), so we report means only.
Configurations that exceeded shared-memory capacity or register limits at high pipeline stages were recorded as launch failures.
\subsection{Evaluation Results}

For each KernelBench problem, we report the output of all three stages of the workflow.
Stage~1 (Pattern Discovery) describes the patterns the agent identified from the computation graph.
Stage~2 (Pattern Realization) presents correctness verification, auto-tuning results, and performance comparisons against the PyTorch cuBLAS baseline.
Stage~3 (Pattern Composition) applies only to \seqsplit{multi-operator} problems, where multiple realized kernels are composed into a single optimized computation graph.

We evaluate CUTLASS GEMM synthesis and auto-tuning (Section~4.2) on three KernelBench Level~1 problems: square GEMM (Problem~1), large-K GEMM (Problem~6), and batched GEMM (Problem~3), presenting results across both Ampere and Hopper architectures that differ significantly in their tunable parameters and schedules.
For multi-pattern composition, we run the full three-stage workflow on two Level~3 blocks: the \mbox{MiniGPT} block from KernelBench problem~\seqsplit{\texttt{44\_MiniGPTBlock}} and the Llama~3 8B decoder block extracted from HuggingFace (Section~5.1).

\subsubsection{KernelBench Problem~1: Square Matrix Multiplication}

In Stage 1,
The agent identified the operation type and tensor dimensions for the square matrix multiplication benchmark and proposed optimization patterns for both Ampere (SM80) and Hopper (SM90) architectures. For brevity, we show only the patterns for the square GEMM in Listing~\ref{lst:proposed_patterns}; the JSON pattern for all patterns follow the same format.

\begin{lstlisting}[caption={Proposed patterns for square matrix multiplication ($4096 \times 4096$). Pattern p1\_ampere (TF32 on Ampere SM80) and p1\_hopper (FP16 on Hopper SM90) are shown.},label={lst:proposed_patterns},basicstyle=\tiny\ttfamily,breaklines=true,breakatwhitespace=false,linewidth=0.92\columnwidth,frame=single,showstringspaces=false]
{
  "pattern_id": "p1_ampere",
  "name": "Ampere TF32 Tensor-Core GEMM",
  "optimization_rule": "GEMM",
  "target_architecture": "SM80 (Ampere)",
  "input_shapes": {"A": [4096,4096], "B": [4096,4096]},
  "data_type": "tf32",
  "implementation_notes": {
    "pipelining": "Multistage cp.async pipeline",
    "grid_schedule": "Data-parallel 2D tiling",
    "tensor_cores": "TF32 tensor cores, inst shape 16x16x16"
  },
  "supporting_example": "cutlass/3.8.0/examples/41_tensorop_ampere_tf32_gemm"
}
{
  "pattern_id": "p1_hopper",
  "name": "Hopper FP16 Tensor-Core GEMM",
  "optimization_rule": "GEMM",
  "target_architecture": "SM90 (Hopper)",
  "input_shapes": {"A": [4096,4096], "B": [4096,4096]},
  "data_type": "fp16",
  "computation_precision": "fp16 with fp32 accumulator",
  "implementation_notes": {
    "pipelining": "Warp-specialized pipeline with TMA",
    "grid_schedule": "Data-parallel 2D tiling",
    "tensor_cores": "FP16 tensor cores, inst shape 16x16x32"
  },
  "supporting_example": "cutlass/examples/48_hopper_warp_specialized_gemm"
}
\end{lstlisting}

Pattern~p1\_ampere targets Ampere's TF32 tensor cores with a multistage software pipeline using \texttt{cp.async} for data prefetching, which is the standard high-performance configuration for GEMM on SM80.
Pattern~p1\_hopper proposes a Hopper warp-specialized design with TMA, targeting the H100.

In Stage 2,
The CUTLASS kernel for pattern~p1\_ampere was synthesized based on CUTLASS example~41 and successfully passed correctness verification with bit-exact agreement against PyTorch's cuBLAS output.
The agent then performed auto-tuning by inferring the search space from the CUTLASS test files for the target architecture and data type, sweeping 98~configurations across 14~tile shapes and pipeline stages 2--8.
Of these, 66~executed successfully and 32~failed due to shared-memory or register overflow at high pipeline stages.

\begin{figure*}[!htbp]
\centering
\includegraphics[width=0.95\textwidth,height=2in]{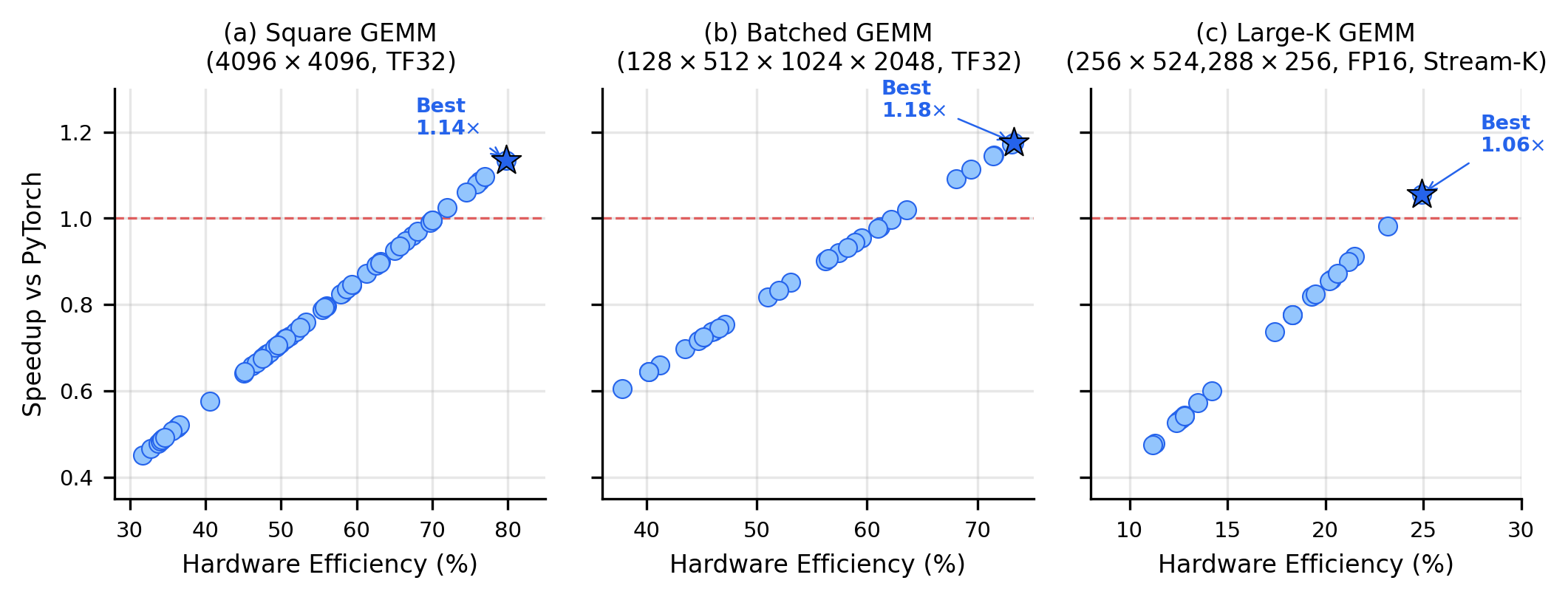}
\caption{Auto-tuning results for all three Level~1 problems on A100. Each point is one configuration; stars mark the best. Dashed lines indicate parity with PyTorch (cuBLAS, TF32). For Problem~6 (c), CUTLASS configurations use mixed precision (FP16 inputs with FP32 compute/output) while the PyTorch baseline uses TF32; the speedup is computed against this TF32 baseline.}
\label{fig:autotune_results}
\end{figure*}

\begin{figure*}[!htbp]
\centering
\includegraphics[width=0.95\textwidth,height=2in]{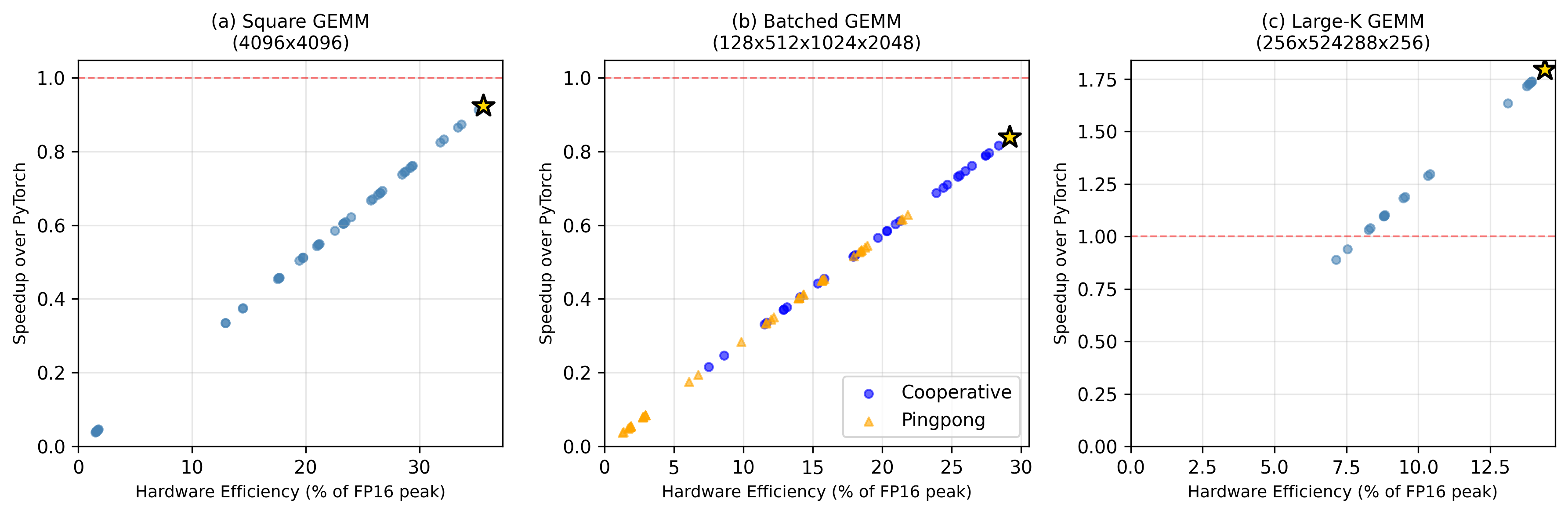}
\caption{Auto-tuning results for all three Level~1 problems on H100. Each point is one configuration; stars mark the best. For Batched GEMM (b), cooperative (blue) and pingpong (orange triangles) schedules are shown separately. All kernels use FP16 inputs with FP32 accumulation. For Large-K GEMM (c), dashed lines indicate parity with PyTorch TF32 baseline (FP16 accumulators overflow for this problem size).}
\label{fig:hopper_autotune}
\end{figure*}

Figure~\ref{fig:autotune_results}~(a) plots all 66~successful configurations as a scatter of hardware efficiency versus speedup over PyTorch.
The best overall configuration---threadblock tile $128 \times 256 \times 32$ with 3~pipeline stages (marked by the star)---achieves 124.4~TFLOPS (79.8\% of peak) and a $1.14\times$ speedup over PyTorch.

For Hopper (SM90), the agent synthesized pattern p1\_hopper (FP16 with FP32 accumulator) based on CUTLASS example~48 and verified correctness.
The agent swept 48 configurations across tile shapes and cluster shapes (cooperative and pingpong schedules).
Figure~\ref{fig:hopper_autotune}~(a) shows the results; the best configuration (tile $128 \times 256 \times 64$, cluster $2 \times 1 \times 1$) achieved 705.0~TFLOPS (35.6\% of FP16 peak) at 0.195~ms, but was $0.92\times$ slower than PyTorch FP16 baseline (0.180~ms), indicating room for further optimization on Hopper.

\subsubsection{KernelBench Problem~3: Batched Matrix Multiplication}

In Stage 1,
The agent examined the operation and input shapes for the batched matrix multiplication benchmark ($A \in \mathbb{R}^{128 \times 512 \times 1024}$, $B \in \mathbb{R}^{128 \times 1024 \times 2048}$) and proposed optimization patterns for both Ampere (SM80) and Hopper (SM90) architectures following the same pattern structure.

Pattern~p1\_ampere targets Ampere's TF32 tensor cores with CUTLASS's \texttt{kBatched} mode for batched workloads.
Pattern~p1\_hopper proposes a Hopper warp-specialized design with TMA and Ptr-Array Batched mode, supporting both cooperative and pingpong schedules.

In Stage 2,
The CUTLASS kernel for pattern~p1\_ampere was synthesized based on CUTLASS example~5 and passed correctness verification for all 30~auto-tuned configurations using.
The agent swept 30~configurations across tile shapes and pipeline stages.
Figure~\ref{fig:autotune_results}~(b) plots the results: configurations with wider N-dimension tiles ($N_{\text{tile}}=256$) consistently outperform narrower variants, and the best configuration achieves 73.3\% of peak.
The best \seqsplit{configuration}---threadblock tile $128 \times 256 \times 32$ with 3~pipeline stages---achieves 114.4~TFLOPS (73.3\% of peak) and a $1.18\times$ speedup over PyTorch cuBLAS.

For Hopper (SM90), the agent synthesized pattern p1\_hopper (FP16 with FP32 accumulator) based on CUTLASS example~56 and verified correctness.
The agent swept 48 configurations across tile shapes and cluster shapes (cooperative and pingpong schedules).
Figure~\ref{fig:hopper_autotune}~(b) shows the results; the best configuration (tile $128 \times 256 \times 64$, cluster $1 \times 1 \times 1$) achieved 577.4~TFLOPS (29.2\% of FP16 peak) at 0.476~ms, compared to cuBLAS FP16 at 0.400~ms ($0.84\times$ speedup).

\subsubsection{KernelBench Problem~6: Large-K Matrix Multiplication}

In Stage 1,
The agent analyzed the operation requirements and dimensions for the large-K matrix multiplication benchmark ($A \in \mathbb{R}^{256 \times 524{,}288}$, $B \in \mathbb{R}^{524{,}288 \times 256}$) and proposed optimization patterns for both Ampere (SM80) and Hopper (SM90) architectures following the same pattern structure.

Pattern~p1\_ampere targets Ampere tensor cores with Stream-K scheduling~\cite{osama2023streamk} for extreme K/M aspect ratios.
Pattern~p1\_hopper proposes a Hopper warp-specialized cooperative Stream-K design with TMA.

In Stage 2,
The CUTLASS kernel for pattern~p1\_ampere was synthesized based on CUTLASS example~47 and verified correctness.
PyTorch FP16 failed due to overflow; the agent configured FP16 for A/B inputs with FP32 accumulator and output to resolve this.
The agent swept 29~configurations across 14~tile shapes and varying pipeline stages.
Figure~\ref{fig:autotune_results}~(c) plots the 23~successful configurations; the best configuration ($64 \times 128 \times 64$, 4 stages) achieves a $1.06\times$ speedup over PyTorch TF32 baseline.

For Hopper (SM90), the agent synthesized pattern p1\_hopper (FP16 with FP32 accumulator) based on CUTLASS example~48 and verified correctness.
Just as with Ampere, PyTorch FP16 overflow occurred in Hopper experiments; using FP32 for both accumulator and output resolved this.
The agent swept 16 configurations across tile shapes and cluster shapes.
Figure~\ref{fig:hopper_autotune}~(c) shows the results; the best configuration (tile $128 \times 128 \times 64$, cluster $2 \times 2 \times 1$) achieved 284.8~TFLOPS (14.4\% of FP16 peak) at 0.241~ms, achieving 1.80$\times$ speedup over PyTorch TF32 baseline (0.433~ms).

On representative GEMM problems (square, batched, and large-K matrix multiply), the agent-synthesized CUTLASS kernels achieved speedups over the cuBLAS baseline after auto-tuning.
On Ampere (A100), speedups ranged from $1.06\times$ to $1.18\times$, with the largest gains on the batched workload where the $128 \times 256 \times 32$ tile shape better exploits non-square output dimensions than cuBLAS's internal heuristics.
For the large-K problem, the agent resolved PyTorch FP16 overflow by using FP16 inputs with FP32 accumulator and output, demonstrating the importance of adaptive precision handling when compiler-generated templates fail.
On Hopper (H100), auto-tuning achieved $0.84\times$--$1.80\times$ speedups across the same representative problems, with large-K GEMM showing the strongest improvement ($1.80\times$) due to effective FP16+FP32 precision handling and stream-K scheduling.

\begin{figure}[!t]
\centering
\includegraphics[width=0.48\textwidth, height=2.3in]{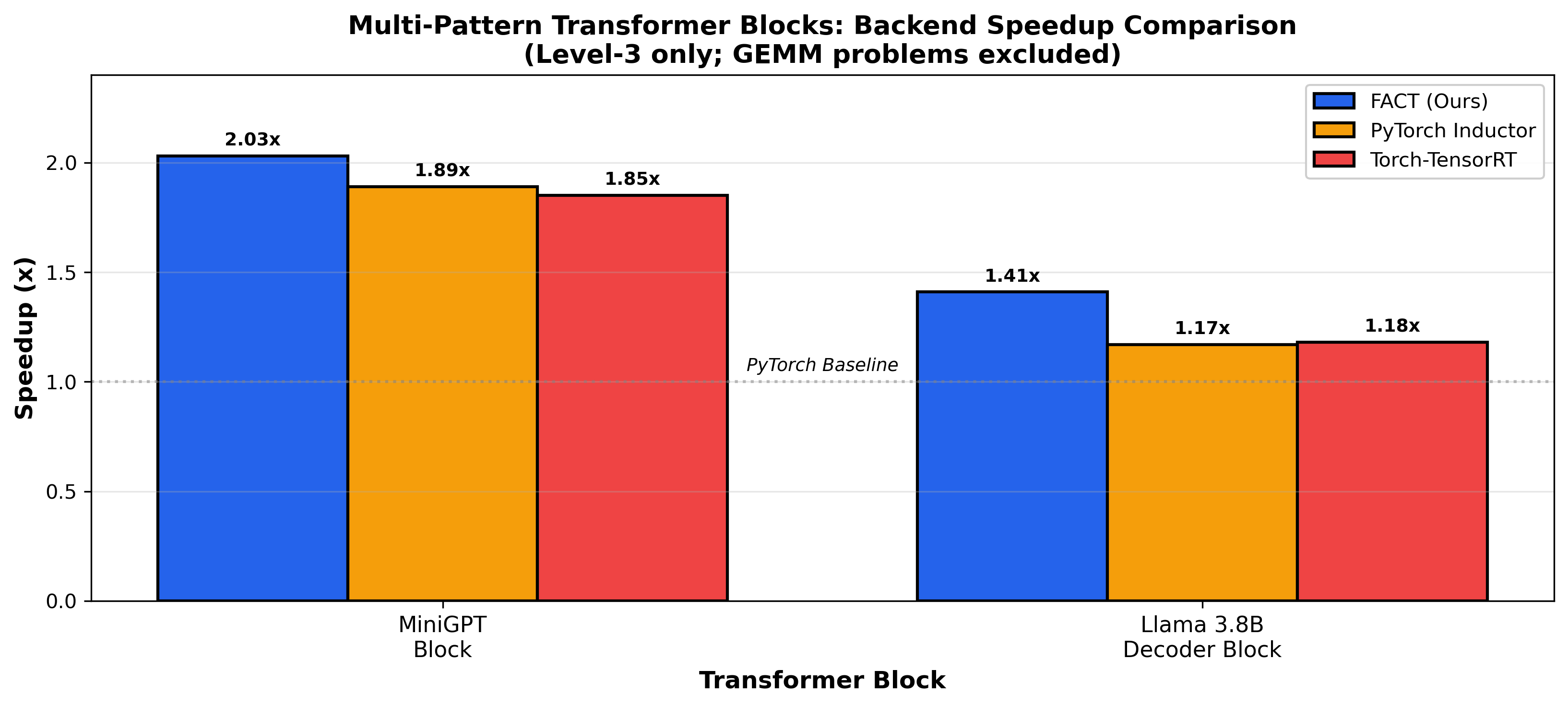}
\caption{Backend speedup comparison on multi-pattern Level~3 transformer blocks: MiniGPT and Llama~3 8B. Bars show FACT, PyTorch Inductor (max-autotune), and Torch-TensorRT (dynamic=False) speedups relative to PyTorch eager baseline. FACT's excels on multi-pattern blocks where subgraph-specific tuning provides substantial gains.}
\label{fig:backend_comparison}
\end{figure}

\subsubsection{KernelBench Level~3: MiniGPT Block (\texttt{44\_MiniGPTBlock})}

We evaluate KernelBench Level~3 problem~\texttt{44\_MiniGPTBlock}: a causal self-attention block paired with a two-layer MLP (expansion to 3072, GELU, projection to 768), at $(B,T,C)=(128,512,768)$ on the same A100 hardware and PyTorch eager baseline as above.

In Stage 1,
The agent analyzed the traced computation graph and identified two high-priority patterns for optimization:

\textbf{Pattern~p1 (FMHA):} Fused Multi-Head Attention over the self-attention subgraph. 
Inputs are FP32, computed in FP16 with FP32 accumulation, and output as FP32.

\textbf{Pattern~p2 (MLP with GELU):} Fuses two consecutive linear transformations with GELU activation in the feedforward network:
\begin{itemize}
  \item GEMM~1: First linear projection (dense layer) with dimensions 768→3072 and GELU activation. FP16 inputs, FP32 accumulation and output.
  \item GEMM~2: Second linear projection (dense layer) with dimensions 3072→768 and standard epilogue. FP16 inputs, FP32 accumulation and output.
\end{itemize}

In Stage 2,
Pattern~p1 was realized as a fused multi-head attention kernel and verified against the PyTorch reference at multiple $(B,T)$ settings.
Pattern~p2 was realized as two fused CUTLASS GEMMs: the first with GELU epilogue fusion (768→3072) and the second with standard epilogue (3072→768), with the auto-tuned configuration passing correctness verification.
Both patterns use mixed-precision kernels with element-wise comparison and tolerance thresholds to ensure numerical accuracy.

\begin{figure}[!t]
\centering
\includegraphics[width=0.48\textwidth, height=2.3in]{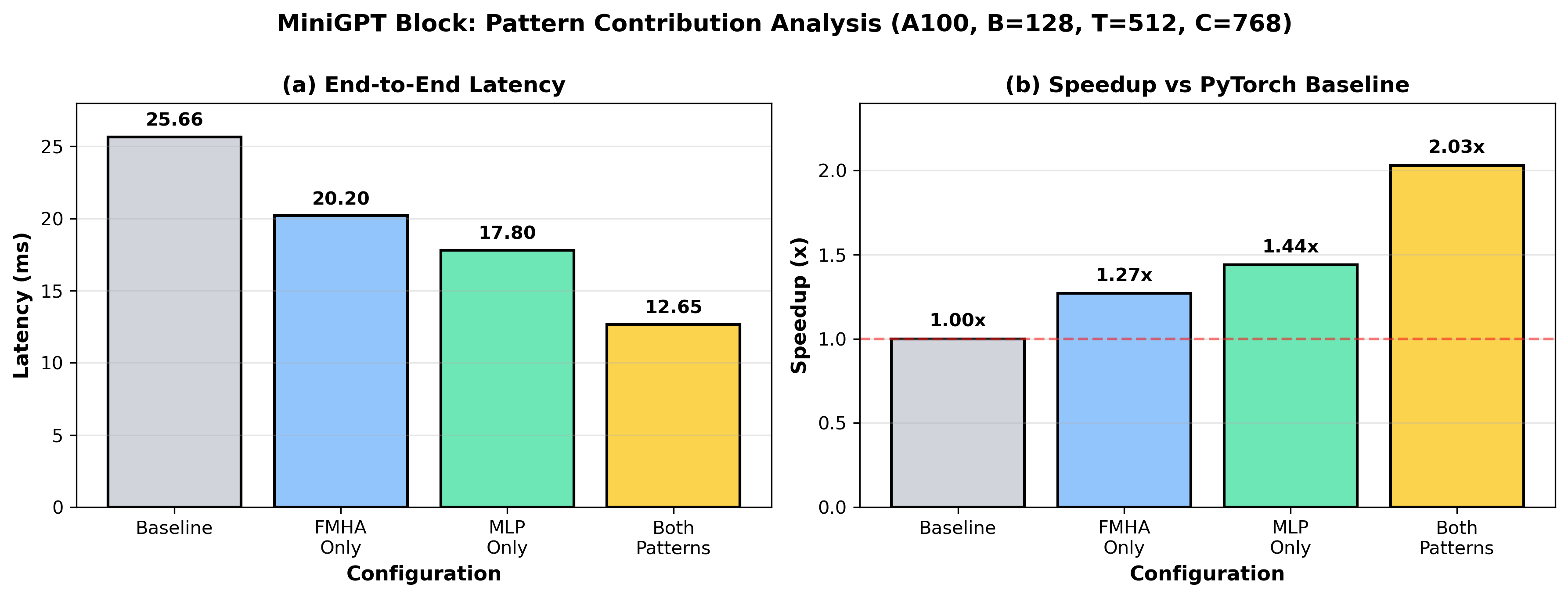}
\caption{KernelBench \texttt{44\_MiniGPTBlock} on A100 with $(B,T,C)=(128,512,768)$. (a)~End-to-end latency for PyTorch baseline, single-pattern optimizations, and both patterns composed. (b)~Speedup relative to the PyTorch baseline measured in the same benchmark run as each optimized configuration (ablations leave the non-target subgraph in eager PyTorch).}
\label{fig:minigpt_block_ablation}
\end{figure}

In Stage 3,
The two extensions were composed into a single \texttt{ModelNew} module, replacing the attention and MLP subgraphs while preserving numerics. End-to-end benchmarks were conducted on the reference shape $(128,512,768)$.
With both patterns enabled, the optimized model achieves $2.03\times$ speedup over the PyTorch baseline (25.665~ms $\rightarrow$ 12.654~ms). Ablation studies reveal: FMHA-only yields $1.27\times$ speedup (20.195~ms), MLP-only yields $1.44\times$ speedup (17.803~ms), and both combined yield $2.03\times$ speedup, as shown in Figure~\ref{fig:minigpt_block_ablation}. These measurements show that the MLP pattern contributes more than the FMHA pattern on this block.

Compiler baselines on the same configuration: PyTorch Inductor (max-autotune) achieves 13.573~ms, and Torch-TensorRT achieves 13.907~ms, both slower than FACT's optimized result. Figure~\ref{fig:backend_comparison} compares FACT against compiler baselines on both transformer blocks.

\subsubsection{Llama~3 8B Decoder Block}

We evaluate a Llama~3 8B decoder block extracted from HuggingFace: a causal self-attention block with Grouped-Query Attention (GQA) paired with a SwiGLU feedforward network, at $(B,T,C)=(16,2048,4096)$ on the same A100 hardware.

In Stage 1,
The agent analyzed the traced computation graph and identified two high-priority patterns:

\textbf{Pattern~p1 (FMHA-GQA):} Grouped-Query Attention with K/V heads (8) fewer than Q heads (32). 
FP32 inputs are expanded via \texttt{repeat\_interleave} before the FMHA kernel call. 
Computation uses FP16 with FP32 accumulation and FP32 output.
Tile parameters: \texttt{kQueriesPerBlock}=32, \texttt{kKeysPerBlock}=128, optimized for \texttt{head\_dim}=128.

\textbf{Pattern~p2 (SwiGLU MLP):} Fuses four operations across two GEMMs:
\begin{itemize}
  \item GEMM~1: \texttt{gate\_proj} (4096→14,336) with SiLU epilogue; \texttt{up\_proj} (4096→14,336) with standard epilogue; elementwise multiply $\text{SiLU}(\text{gate}) \times \text{up}$. FP16 inputs, FP32 accumulation and output.
  \item GEMM~2: \texttt{down\_proj} (14,336→4096) with standard epilogue. FP16 inputs, FP32 accumulation and output.
\end{itemize}

In Stage 2,
Pattern~p1 was realized as a fused multi-head attention kernel with GQA support and verified against the PyTorch reference across multiple input shapes.
Pattern~p2 was realized as two fused CUTLASS kernels: the first combining gate projection with SiLU activation and up projection, and the second implementing down projection.
Both patterns use mixed-precision kernels verified using the method described in Section~\ref{sec:experimental_setup}.

\begin{figure}[!t]
\centering
\includegraphics[width=0.48\textwidth]{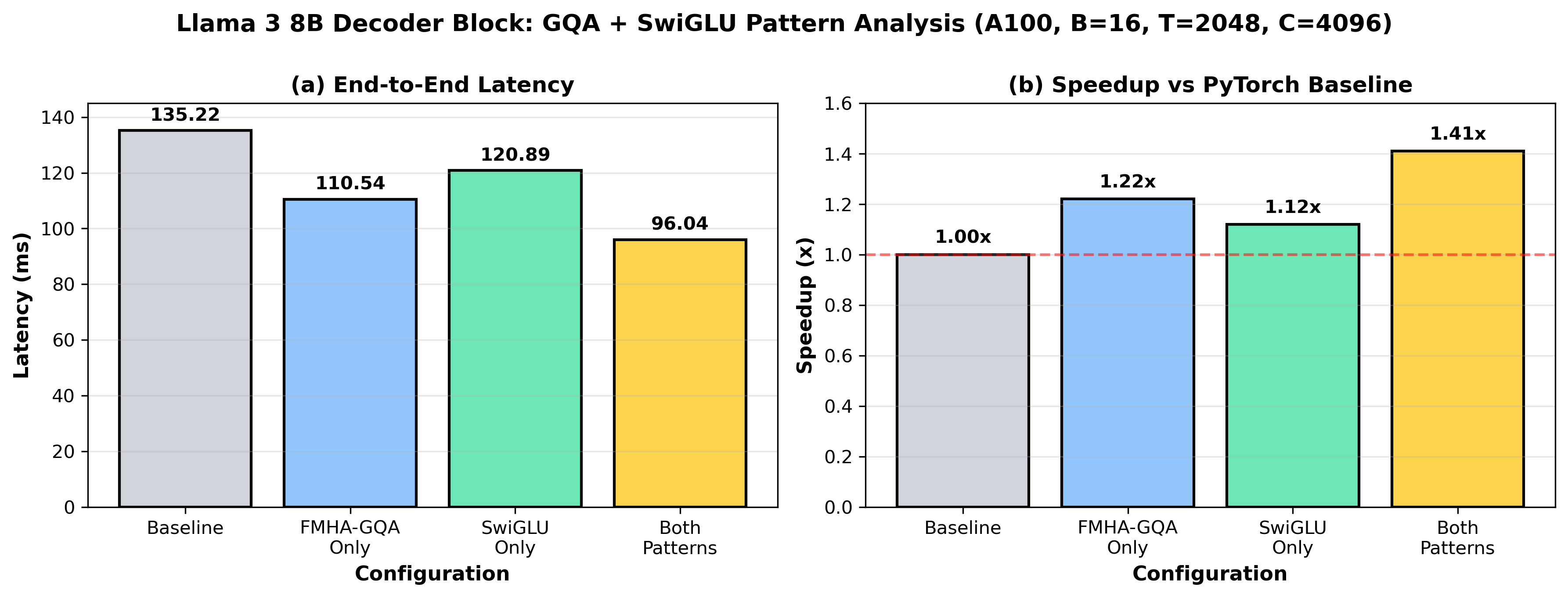}
\caption{Llama~3 8B decoder block on A100 with $(B,T,C)=(16,2048,4096)$. (a)~End-to-end latency for PyTorch baseline, single-pattern optimizations (FMHA-GQA and SwiGLU), and both patterns composed. (b)~Speedup relative to the PyTorch baseline measured in the same benchmark run as each optimized configuration.}
\label{fig:llama_block_ablation}
\end{figure}

In Stage 3,
The two extensions were composed into a single \texttt{ModelNew} module, replacing the attention and MLP subgraphs. End-to-end benchmarks were conducted on the reference shape $(16,2048,4096)$.
With both patterns enabled, the optimized model achieves $1.41\times$ speedup over the PyTorch baseline (135.215~ms $\rightarrow$ 96.038~ms). Ablation studies reveal: FMHA-GQA-only yields $1.22\times$ speedup (110.541~ms), SwiGLU-only yields $1.12\times$ speedup (120.887~ms), and both combined yield $1.41\times$ speedup, as shown in Figure~\ref{fig:llama_block_ablation}. The FMHA-GQA pattern contributes more substantial improvement than SwiGLU on this larger, more attention-heavy block.

Compiler baselines on the same configuration: PyTorch Inductor (max-autotune) achieves 115.741~ms, and Torch-TensorRT achieves 114.133~ms, both substantially slower than FACT's optimized result and showing only marginal speedup over eager PyTorch. Figure~\ref{fig:backend_comparison} compares FACT against compiler baselines on both transformer blocks.

\subsection{Discussion}

\textbf{Key findings.}
Our evaluation demonstrates that library-grounded agentic kernel synthesis can achieve competitive performance on both single-operator and multi-pattern workloads.
On \seqsplit{Level-1} GEMM problems, auto-tuned CUTLASS kernels achieved 1.06$\times$--1.18$\times$ speedups over PyTorch's cuBLAS baseline, with the strongest improvement (1.18$\times$) on batched GEMM where cuBLAS's heuristics are less tuned for irregular shapes.
On Level-3 MiniGPT block, composing fused multi-head attention with GELU-fused MLP yields $2.03\times$ end-to-end speedup, with MLP contributing more (1.44$\times$) than FMHA (1.27$\times$).
On the larger Llama~3 8B block with GQA and SwiGLU, composing FMHA-GQA with SwiGLU MLP yields $1.41\times$ end-to-end speedup, with FMHA-GQA contributing more (1.22$\times$) than SwiGLU (1.12$\times$). These results demonstrate that the relative contribution of patterns depends on workload characteristics: smaller blocks benefit more from MLP optimization, while larger attention-heavy models benefit more from attention optimization.

\textbf{Compiler baseline comparison.}
We evaluated PyTorch Inductor and Torch-TensorRT on multi-pattern transformer blocks (MiniGPT and Llama~3 8B), excluding Level-1 GEMM problems from compiler comparison as they represent isolated single-operator workloads rather than multi-pattern composition targets.
For \seqsplit{Mini-GPT}, Inductor achieves 1.89$\times$ and Torch-TensorRT achieves 1.85$\times$ speedups---both lower than FACT's 2.03$\times$.
For Llama~3 8B, Inductor achieves 1.17$\times$ and Torch-TensorRT achieves 1.18$\times$ speedups---both lower than FACT's 1.41$\times$.
Our results suggest that multi-pattern composition with library-grounded kernels can outperform general-purpose compilers when workload-specific patterns are known.

\section{Related Work}
\label{sec:related-work}

\textbf{Agentic GPU kernel generation and optimization.}
Recent work demonstrates that LLM-driven agents can generate and optimize CUDA kernels by closing the loop with compilation, correctness checking, and profiling.
KernelBlaster proposes memory-augmented in-context RL to transfer optimization knowledge across tasks and improve KernelBench speedups~\cite{kernelblaster_arxiv2026}.
CUDA Agent scales agentic RL with automated verification/profiling to learn high-performance CUDA kernel generation behavior~\cite{cuda_agent_arxiv2026}.
StitchCUDA extends this direction to end-to-end GPU programs with a planner/coder/verifier decomposition and rubric-based RL, emphasizing robust rewards from real executions~\cite{stitchcuda_arxiv2026}.
Astra studies a multi-agent workflow that starts from existing CUDA kernels (rather than purely from high-level specs) and iteratively applies profiling-driven transformations to improve performance~\cite{astra_arxiv2025}.

\textbf{LLM-guided optimization for accelerators and compilers.}
Autocomp targets tensor accelerators and frames optimization as an LLM-driven search guided by domain knowledge and hardware feedback, and shows that optimization schedules can be reused across related tensor operations~\cite{autocomp_mlarchsys2025}.

\textbf{Subgraph-aware auto-tuning.}
Closest to our dynamic pattern registry is prior work on exploiting subgraph similarities for efficient auto-tuning~\cite{li2023familyseer}, which introduces FamilySeer, a framework that caches and reuses tuning results across similar tensor program subgraphs.
Our work differs in three key respects: (1) we use LLM agents to discover patterns automatically rather than relying on manual subgraph extraction; (2) our pattern registry is indexed by architecture, data type, and shape in addition to subgraph structure; and (3) we focus on CUTLASS template instantiation rather than search-space exploration within a single compiler backend.

\textbf{Library-centric execution engines and tuning.}
CUrator focuses on efficient LLM execution by integrating CUDA libraries (cuBLAS/CUTLASS) through compiler infrastructure and profiling-driven selection of tiling and split-K strategies for GEMM variants~\cite{curator_cgo2025}.
Collectively, these systems motivate our focus on expanding the search space to newer CUTLASS 3.x kernel schedules (e.g., warp-specialized cooperative and pingpong) and enabling agent-driven exploration for Hopper-specific performance regimes.

\section{Conclusion and Future Work}

We presented FACT, a framework that combines two key contributions: (1) GEMM synthesis with architecture-aware auto-tuning to optimize matrix multiplication kernels, and (2) compositional kernel synthesis with a dynamic pattern registry to optimize complex multi-operator patterns.
Our three-stage workflow---Pattern Discovery, Pattern Realization, and Pattern Composition---enables automated discovery and synthesis of optimized kernel patterns.
On square, large-K, and batched GEMM problems with input shapes from KernelBench Level-1, auto-tuned CUTLASS kernels achieved speedups ranging from 1.06$\times$--1.18$\times$ on Ampere A100 and 0.84$\times$--1.80$\times$ performance variations on H100 over cuBLAS.
On Level-3 transformer blocks, FACT achieves 2.03$\times$ end-to-end speedup on MiniGPT and 1.41$\times$ on Llama~3 8B, consistently outperforming compiler baselines (Inductor: 1.89$\times$ and 1.17$\times$ respectively; TensorRT: 1.85$\times$ and 1.18$\times$).

\textbf{Limitations and Future Work.}

Our framework currently relies on the CUTLASS example index as a foundation for pattern synthesis. While the example catalog is extensive, patterns that significantly deviate from existing examples may require the agent to synthesize entirely new kernels from scratch---a scenario we have not yet explored. We plan to investigate this capability in future work to understand the boundaries of library-grounded synthesis versus de novo kernel generation.

Additionally, FACT is currently scoped to CUTLASS, which limits its applicability to NVIDIA GPUs. We are planning to extend FACT to incorporate multiple kernel generation DSLs, including Triton and CuTe. We believe our agentic workflow and pattern registry design are sufficiently generalizable that extending to new DSLs will not require fundamental rearchitecting of the framework.

Finally, FACT targets GPU acceleration only. A significant frontier lies in supporting diverse accelerators such as Google TPUs and NPUs, which, despite their computational power, lack mature programming ecosystems comparable to NVIDIA's. We are working toward extending FACT to these accelerators, which would enable broader adoption of kernel optimization techniques across heterogeneous hardware platforms.


\bibliographystyle{ACM-Reference-Format}
\bibliography{references}

\end{document}